\pdfoutput=1
\documentclass[prl,twocolumn,showpacs,amsmath,amssymb,floatfix,superscriptaddress]{revtex4}
\usepackage{graphicx}
\usepackage{dcolumn}
\usepackage{bm}
\usepackage{epsf}
\begin{document}

\title{Nonperturbative Quantum Physics from Low-Order Perturbation Theory}

\author{H\'ector Mera}
\email{hmera@udel.edu}
\affiliation{Department of Physics and Astronomy, University of Delaware, Newark, DE 19716-2570, USA}
\author{Thomas G. Pedersen}
\affiliation{Department of Physics and Nanotechnology, Aalborg University, DK-9220 Aalborg East, Denmark}
\author{Branislav K. Nikoli\'c}
\affiliation{Department of Physics and Astronomy, University of Delaware, Newark, DE 19716-2570, USA}

\begin{abstract}
The Stark effect in hydrogen and the cubic anharmonic oscillator furnish examples of quantum systems where the perturbation results in a certain ionization probability by tunneling processes. Accordingly, the perturbed ground-state energy is shifted and  broadened, thus acquiring an imaginary part which is considered to be a paradigm of nonperturbative behavior. Here we demonstrate how the low order coefficients of a divergent perturbation series can be used to obtain excellent approximations to both real and imaginary parts of the perturbed ground state eigenenergy.  The key is to use analytic continuation functions with a built in analytic structure within the complex plane of the coupling constant, which is tailored by means of Bender-Wu dispersion relations. In the examples discussed the analytic continuation functions are Gauss hypergeometric functions, which take as input fourth order perturbation theory and return excellent approximations to the complex perturbed eigenvalue.  These functions are  Borel-consistent and dramatically outperform widely used Pad\'e and Borel-Pad\'e approaches, even for rather large values of the coupling constant.
\end{abstract}


\maketitle

Since the pioneering work of Dyson~\cite{Dyson1952}, who provided an argument as to why perturbation series in quantum electrodynamics (QED) are divergent, the fundamental problem of how to reconstruct physical observables from power-series expansions with zero convergence radius has remained an active area of research~\cite{Caliceti2007,Marino2014}. This problem  has been encountered in virtually all areas of quantum physics, such as  statistical~\cite{Brezin1977,Kleinert2001,Yukalov1990}, string~\cite{Marino2014} and quantum field theories~\cite{Lipatov1977,Dunne2002}, as well as many-body problems of condensed matter physics~\cite{Pollet2010}.

The simplest examples can be found  in single-particle quantum mechanics~\cite{Caliceti2007,Marino2014,Brezin1977,Dunne2002,Bender1969,Janke1995,Jentschura2001}.  For instance,  the perturbation expansion for the Stark Hamiltonian has zero radius of convergence~\cite{Jentschura2001,Privman1980}. When a hydrogen atom is placed in a homogeneous electric field, the electronic ground state energy is shifted and broadened as the electric field intensity, $F$, increases. As a function of $F$, the perturbed ground state energy then acquires both a real part $\Delta$ and an imaginary part $\Gamma/2$, $E(F) = \Delta(F) - i\Gamma(F)/2$. The latter reflects the tunneling rate in and out of the Coulomb potential, which  is very difficult to obtain perturbatively. To see this, let $f \equiv(F/4)^2$ and consider the perturbative expansion~\cite{Privman1980} for the ground state energy of hydrogen in  powers of $F$ around $F=0$,
\begin{eqnarray}
E(F) & \sim & -\frac{1}{2}\sum_{n=0}^\infty e_n f^n \label{eq:asym} \\
& = & -\frac{1}{2}\left(1+72 f+28\,440 f^2+40\,204\,464f^3+\cdots\right),\nonumber
\end{eqnarray}
where atomic units (a.u.) are used throughout the paper. The same-sign expansion coefficients $e_n$ are real and grow  factorially with $n$. Therefore, the series in Eq.~(\ref{eq:asym}) has zero radius of convergence. No matter how small $F$  is,  the series in Eq.(\ref{eq:asym}) will never converge to $E(F)$, and so ``$\sim$'' is used in Eq.~(\ref{eq:asym}) to indicate that the RHS is an asymptotic expansion of the LHS.

Accurate calculations of $\Gamma(F)$ have been achieved by a combination of perturbation theory with Borel-Pad\'e~\cite{Reinhardt1982,Jentschura2001} resummation techniques. Unfortunately, these techniques are ultimately impractical since they require too many coefficients. In most quantum-mechanical problems of interest, one has available only low-order coefficients, obtained  by calculating Feynman diagrams up to a given order and/or by summing a few classes of diagrams (out of infinitely many possible) to ``infinite order''~\cite{Stefanucci2013,Mera2013}. In contrast, the recent development of the diagrammatic Monte Carlo (DiagMC) method~\cite{Pollet2010} has made it possible to sample Feynman diagrams up to large orders. However, DiagMC is computationally demanding and relies on resummation and regularization techniques~\cite{Pollet2010} in order to treat divergent or slowly convergent series. Other resummation approaches~\cite{Kleinert2001} rely on large-order information~\cite{Bender1969,Brezin1977} and knowledge of $E(F)$ for large values of $F$ to design rapidly converging strong-coupling expansions from divergent weak-coupling expansions. While promising, these approaches have remained largely unexplored.

These considerations lead us to the following questions:  Do we really need a large number of coefficients to predict quantities like $\Gamma(F)$? Or can it be done with just a few coefficients? In this Letter, we demonstrate that low-order approximations can, paradoxically,  reproduce nonperturbative quantities like $\Gamma(F)$ with excellent accuracy even for rather large values of the perturbation strength. This is achieved by using the so-called Bender-Wu dispersion relations~\cite{Bender1969} to guess the branch cut structure of $E(F)$, for complex $F$. Then we design approximants with the property of having the desired branch-cut structure ``built-in'', with flexible branch points. This flexibility is essential to approximate $E(F)$ in a wide variety of problems. For the Stark problem the branch cut structure is indeed known from Bender-Wu dispersion relations \cite{Bender1969}: $E(F)$ possesses branch points at $F=0$  and  $F\rightarrow \pm \infty$.  

Let us then start by trying to calculate $E(F)$ using a divergent series such as Eq.~\eqref{eq:asym}. Traditionally, the first choice is to calculate Pad\'e approximants~\cite{Baker1996}. These are parameterized rational approximations, $E(F) \approx E_{L/M}(F)$,  of the form
\begin{equation}
E_{L/M}(F)=\frac{\sum_{n=0}^L p_n f^n}{1+\sum_{n=1}^M q_n f^n},
\end{equation}
where the parameters $p_n$ and $q_n$ are determined by equating each order up to $L+M=N$ in the Taylor and asymptotic series of $E_{L/M}(F)$ and $E(F)$, respectively, around $F=0$, so that $E(F)=E_{L/M}(F)+\mathcal{O}(f^{L+M+1})$. Pad\'e approximants and other similarly simple sequence transformations are
 valuable tools for analytic continuation (AC), and can work well in many cases of practical interest~\cite{Mera2013}. They provide a family of rational functions that are easily built order by order: first-order perturbation theory gives $E_{1/0}$ and $E_{0/1}$; second-order perturbation theory yields $E_{2/0}$, $E_{0/2}$ and $E_{1/1}$; etc. By studying the resulting Pad\'e table, one can in many cases  extract  good approximations to the expectation value of interest. However, by approximating $E(F)$ with a rational function of $F$, one is imposing an asymptotic behavior for large values of $F$  which is in general not  physical. Approximating $\Delta(F)$ can be difficult because the denominator in $E_{L/M}$ can vanish for specific values of the interaction strength. More importantly, $E_{L/M}(F)$ is a real number for real $F$, and therefore $\Gamma_{L/M}(F)=0$. This means that the standard Pad\'e approximants cannot work for our problem as they fail to give $\Gamma(F) \neq 0$~\cite{Reinhardt1982,Jentschura2001}.

Nevertheless, the idea behind Pad\'e approximants is very general and it can be used to propose new approximations. For example, one can choose a parameterized analytic function $\mathcal{E}(F)=\mathcal{E}(\{h_i\};F)$ to approximate $E(F)$, fixing the parameters $\{h_i\}$ so that the Taylor series for $\mathcal{E}(F)$ is equal to the asymptotic series of $E(F)$ up to the desired order. An example is provided in Ref.~\cite{Bender1996}, which considers continued-exponential approximations of the form \mbox{$\exp{( h_1 x\exp(h_2 x\exp(h_3 x\ldots)))}$}, where the parameters $h_i$ are fixed from the perturbation expansion of $E(F)$, as is the case with Pad\'e approximants. Since here we are concerned with the determination of $\Gamma(F)$, we initially aimed for a function $\mathcal{E}(F)$ with the following desirable properties: ({\em i}) it is a complex function of real $F$, with the ability to mimic the branch cut structure discussed above; ({\em ii}) it can be built from low-order perturbation theory, as Pad\'e approximants and continued-functions are built; ({\em iii}) it is amenable to generalization by being a member of a more general family of ``higher order'' functions; and ({\em iv}) it is sufficiently general and flexible in order to include many possible functions as particular cases.

A possible candidate satisfying all desirable properties ({\em i})-({\em iv}) is the Gauss hypergeometric function $_2 F_1(h_1,h_2,h_3;h_4 f)$. It satisfies ({\em i}) and ({\em ii}) as it is complex (and has a branch cut) for $h_4 f > 1$, and it contains at most four parameters so it can be built from the coefficients $e_{1}, \ldots, e_4$. It also satisfies condition ({\em iii}) because $_2 F_1$ can be obtained by approximating the ratio between consecutive expansion coefficients, $e_n/e_{n-1}$, by a 1/1 Pad\'e approximant that reproduces exactly $e_n/e_{n-1}$ for $0 < n \le 4$. If higher-order Pad\'e approximants could be used as well, one would obtain hypergeometric functions of higher order, $_p F_q$, which are actually instances of an even ``higher-order'' function---the so-called Meijer-$G$ function~\cite{nistlibrary}. Finally, $_2F_1$ satisfies condition ({\em iv}) as it is well known that many functions are particular cases of $_2 F_1$.

\begin{figure*}[ht!]
\includegraphics[width=\textwidth]{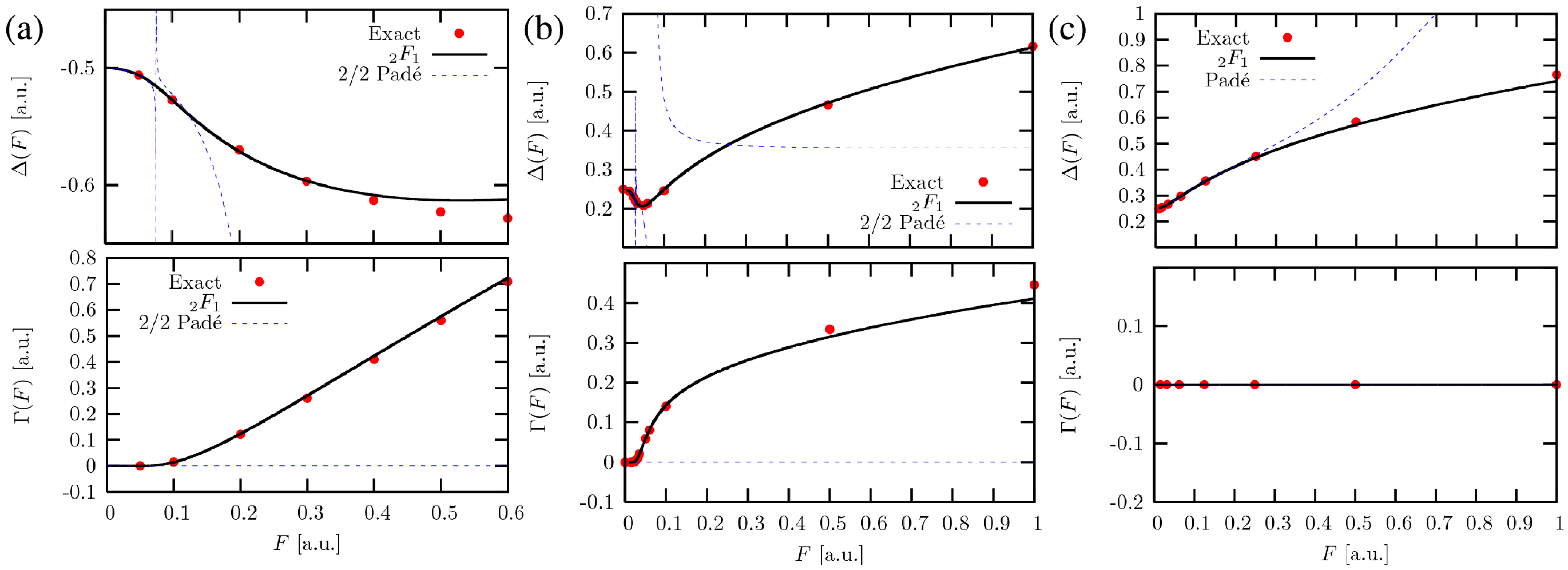}
\caption{(Color online) Real $\Delta(F)$ and imaginary $\Gamma(F)$ part of the perturbed ground state energy of: (a) the Stark Hamiltonian as a function of the electric field strength $F$; (b) the anharmonic oscillator with real perturbation $Fx^3$; and (c) the anharmonic oscillator with imaginary perturbation $iFx^3$. We show a comparison between  numerically exact values taken from the literature (dots)~\cite{Kolosov1987,Alvarez1988,Bender1999}, the fourth-order hypergeometric approximant $_2 F_1$ (solid line) and Pad\'e approximants (dashed line). In all three cases the $_2 F_1$ approximant improves over  Pad\'e approximants [of the same-order in panels (a) and (b); and of much higher order in panel (c)] for the calculation of both $\Delta(F)$ and  $\Gamma(F)$.}
\label{fig:fig1}
\end{figure*}
The Taylor series for $_2 F_1$ is given by:
\begin{equation}
_2 F_1(h_1,h_2,h_3;h_4 f)= \sum_{n=0}^\infty \frac{(h_1)_n(h_2)_n}{n! (h_3)_n} h_4^n f^n,
\end{equation}
where $(h_i)_n=h_i (h_i+1)\cdots(h_i+n-1)$ is a so-called Pochhammer symbol. To obtain the $h_i$, one equates each order in the asymptotic series for $E(F)$  with the corresponding order in the Taylor series for $\mathcal{E}(F)$ to obtain a system of four equations with four unknowns
\begin{equation}
e_n= \frac{(h_1)_n (h_2)_n h_4^n}{(h_3)_n n!},\, 1\le n\le 4.
\end{equation}
Once the coefficients $h_i$ are determined, a hypergeometric approximation $E(F)\approx\mathcal{E}(F)$ for, e.g., the Stark case can be constructed as:
\begin{equation}
\mathcal{E}(F) = -\frac{1}{2} \, _2F_1(h_1,h_2,h_3,h_4 f).
\label{eq:2f1}
\end{equation}
We apply this scheme to three Hamiltonians from single-particle quantum mechanics, offering illustrative examples of divergent perturbation series as well as nonperturbative behavior: the Stark Hamiltonian, $\hat{H}=-\nabla^2/2-1/r+Fz$, with asymptotic series expansion described~\cite{Privman1980,Jentschura2001} by Eq.~\eqref{eq:asym}; the cubic one-dimensional anharmonic oscillator with real perturbation~\cite{Alvarez1988}, $\hat{H}=-(\partial^2/\partial x^2)/2+\lambda x^2/2+Fx^3$; and the cubic one-dimensional anharmonic oscillator with  imaginary perturbation~\cite{Bender1999},  $\hat{H}=-(\partial^2/\partial x^2)/2+\lambda x^2/2 + i Fx^3$. Here, $\lambda$ is the force constant taken as $1/4$ in the numerical analysis below. Furthermore, the perturbed ground-state eigenvalue has $\Gamma(F) \neq 0$ in the first two cases, while in the third case one has a non-Hermitian, but $\mathcal{P}\mathcal{T}$-symmetric~\cite{Bender1999}, Hamiltonian with real eigenvalues. For simplicity, all equations are written  assuming the Stark Hamiltonian problem. 

Figure~\ref{fig:fig1} shows $\Delta(F)$ [top panels] and $\Gamma(F)$ [bottom panels] as a function of $F$ for these three Hamiltonians.  In Fig.~\ref{fig:fig1}(a), values of  $\Delta(F)$ and $\Gamma(F)$ are shown for the Stark Hamiltonian. Exact results taken  from Ref.~\cite{Kolosov1987} are compared with those calculated using the simple hypergeometric approximant $_2F_1$ and  the  same-order 2/2 Pad\'e approximant. The simple $_2F_1$ approximant introduced here provides excellent approximations to both $\Delta(F)$ and $\Gamma(F)$, while the 2/2 Pad\'e approximant fails to approximate either quantity.  In Fig.~\ref{fig:fig1}(b) a similar comparison is made for the cubic anharmonic oscillator with real perturbation, taking the exact numerical values from Ref.~\cite{Alvarez1988}. Once again the $_2F_1$ approximant dramatically outperforms the 2/2 Pad\'e approximant. Finally, in Fig.~\ref{fig:fig1}(c) we see the results obtained from the cubic anharmonic oscillator with imaginary perturbation. In this case, both Pad\'e and exact results are taken from Ref.~\cite{Bender1999}. The Pad\'e results have been obtained in Ref.~\cite{Bender1999} by means of a Cesaro sum of the energies obtained from the 22/22 and 22/23 Pad\'e approximants. Figure~\ref{fig:fig1}(c) shows that $_2 F_1$ outperforms large-order  Pad\'e ($N=44$) for the calculation of $\Delta(F)$, and they both reproduce the exact value of $\Gamma(F)=0$. Therefore the hypergeometric approximant offers an excellent fourth-order approximation,  likely to outperform Pad\'e approximants of much higher order.  Note that a single hypergeometric approximant yields  the results shown in Fig.~1(b) and 1(c), just by replacing $F$ by $\textrm{i}F$.
\begin{figure*}[ht!]
\includegraphics[width=\textwidth]{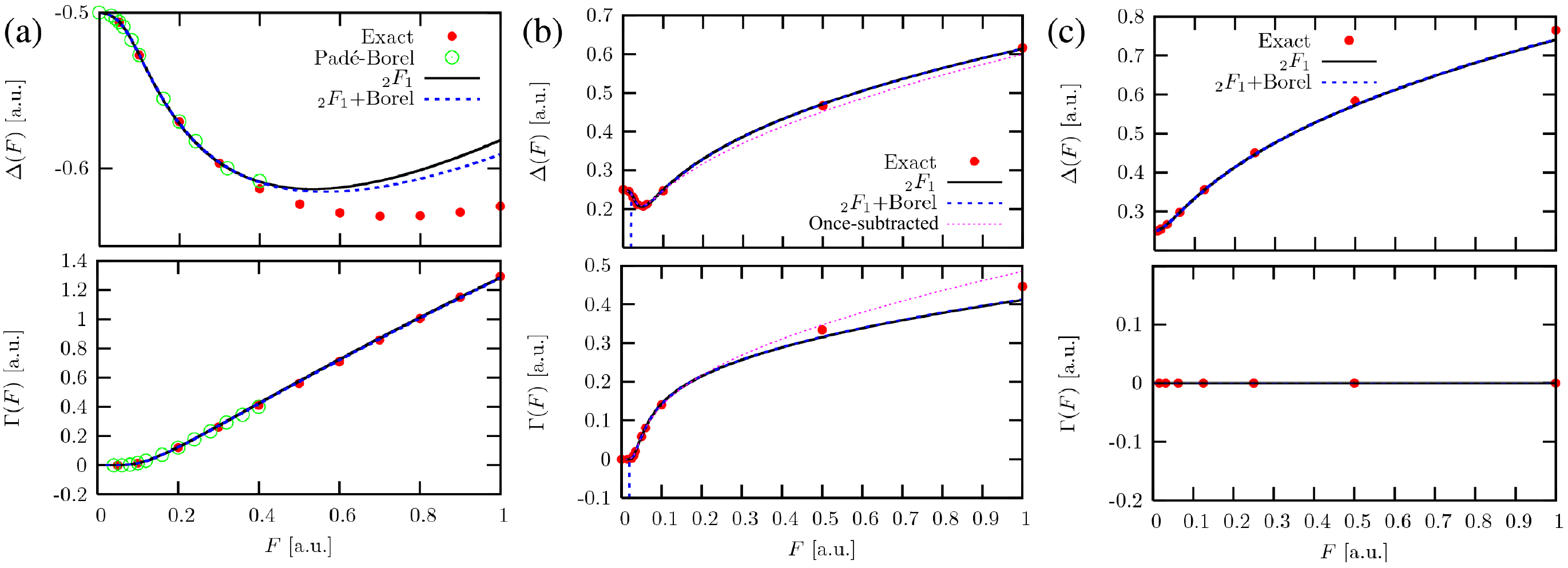}
\caption{(Color online) The same quantities as in Fig.~\ref{fig:fig1}, but calculated using the fourth-order hypergeometric approximant $_2 F_1$ (solid line), Borel-hypergeometric method (dashed line) in Eq.~\eqref{eq:bh} and the numerically exact values taken from the literature (dots)~\cite{Kolosov1987,Alvarez1988,Bender1999}. The Borel-hypergeometric and hypergeometric methods are in excellent agreement, both yielding excellent approximations to both $\Delta(F)$ and $\Gamma(F)$ in all three cases.}
\label{fig:fig2}
\end{figure*} 

A comparison between the hypergeometric approximant and Pad\'e approximants is admittedly not very fair. To
obtain $\Gamma(F) \neq 0$ from Pad\'e approximants, the standard procedure~\cite{Jentschura2001} thus far has been to employ the Borel-Pad\'e method~\cite{Caliceti2007}. In this method, one starts from a large number of coefficients $e_n$ and evaluates the Borel-transformed coefficients $b_n=e_n/n!$, which are then employed to calculate Pad\'e approximants $B_{L/M}(f)$ and integrals
\begin{equation}
\mathcal{F}(f)=\int_0^\infty \,dt B_{L/M}(f t) e^{-t}.
\label{eq:borel1}
\end{equation}
This yields the Borel-Pad\'e  approximation,  \mbox{$E(F)\approx -\frac{1}{2} \mathcal{F}(f)$}. The Borel method  removes $n!$ from the coefficients, sums the transformed series, and puts $n!$ back into the series by means of the Laplace transform in Eq.~\eqref{eq:borel1}. The essence of the Borel-Pad\'e method~\cite{Caliceti2007} is to perform AC on the Borel transformed coefficients and use the resulting analytic function to evaluate the integral in Eq.~(\ref{eq:borel1}). While the Borel-Pad\'e method allows accurate calculations of $\Gamma(F)$  from the perturbation series, it also requires~\cite{Jentschura2001} very large orders of perturbation theory that are unavailable in practice and lead to accuracy issues,  impeding the calculation at high values of the perturbation strength. 

In the Borel-Pad\'e method, the analytic function is a Pad\'e approximant. In the same spirit, we use hypergeometric functions as analytic functions to construct the Borel-hypergeometric method, by performing hypergeometric AC of the Borel-transformed series, calculating the $h_i$ coefficients that define the hypergeometric function $_2 F_1(h_1, h_2, h_3; h_4 f)$ from $e_n/n!$. The Borel-hypergeometric approximation, $E(F)\approx \mathcal{E}(F)$, is then
\begin{equation}\label{eq:bh}
\mathcal{E}(F) \approx -\frac{1}{2} \int_0^\infty dt \, e^{-\alpha t}\,_2 F_1(h_1, h_2, h_3, \alpha h_4 f t),
\end{equation}
and $\alpha=\sqrt{i}$ specifies the integration contour~\cite{Jentschura2001}. An expression somewhat similar to Eq.~\eqref{eq:bh} was used in Ref.~\cite{Kleinert2001} as the starting point to construct convergent strong-coupling expansions, while requiring the knowledge of both $e_{n\rightarrow\infty}$ and $E(F\rightarrow \infty)$.

We now apply the Borel-hypergeometric  method to approximate $\Delta(F)$ and $\Gamma(F)$ for the same three Hamiltonians studied in Fig.~\ref{fig:fig1}.  Figure~\ref{fig:fig2} demonstrates that in all three problems considered the Borel-hypergeometric method gives excellent approximations to both $\Delta(F)$ and $\Gamma(F)$, and reproduces the results given by the hypergeometric approximant. Comparing Borel-hypergeometric and hypergeometric approximants reveals that the hypergeometric approximant is \emph{Borel consistent} to a very good approximation. It is well known that a convergent sum and its Borel resummation give the same result. The hypergeometric approximations discussed here clearly satisfy this highly desirable property. The careful reader will surely notice that in Fig.~2(b) the Borel-hypergeometric sum diverges for very small $F$. This is not a problem since the simple hypergeometric approximant is already well-behaved  for $F\rightarrow0$. Alternatively, one can calculate one extra order of perturbation theory and build the Borel-hypergeometric approximant from the coefficients of the once-subtracted series, $[E(F)-E(0)]/(e_1 f)$. As shown in Fig.~2(b) that procedure mitigates this minor problem, while leading to similarly accurate overall results.  We emphasize that the hypergeometric and Borel-hypergeometric approaches are fourth-order approximations and thus much simpler and less expensive than the widely used Borel-Pad\'e method~\cite{Caliceti2007}, while being of comparable accuracy. For instance, in the case of the Stark Hamiltonian with very large $F=0.4$,  approximately 70 orders of perturbation theory were required in Ref.~\cite{Jentschura2001} as the input for the Borel-Pad\'e scheme to produce  $E(F=0.4)=-0.608 - 0.200i$, which can be contrasted with our result $\mathcal{E}(F=0.4) = -0.609 - 0.212i$, and with numerically exact data \cite{Kolosov1987} $\mathcal{E}(F=0.4) = -0.613 - 0.205i$. For the Stark Hamiltonian, $F=0.4$ a.u. $\simeq 2 \times 10^3$ MV cm$^{-1}$ corresponds to a rather large electric field.

\begin{figure}[t!]
\centering
\includegraphics[width=0.45\textwidth]{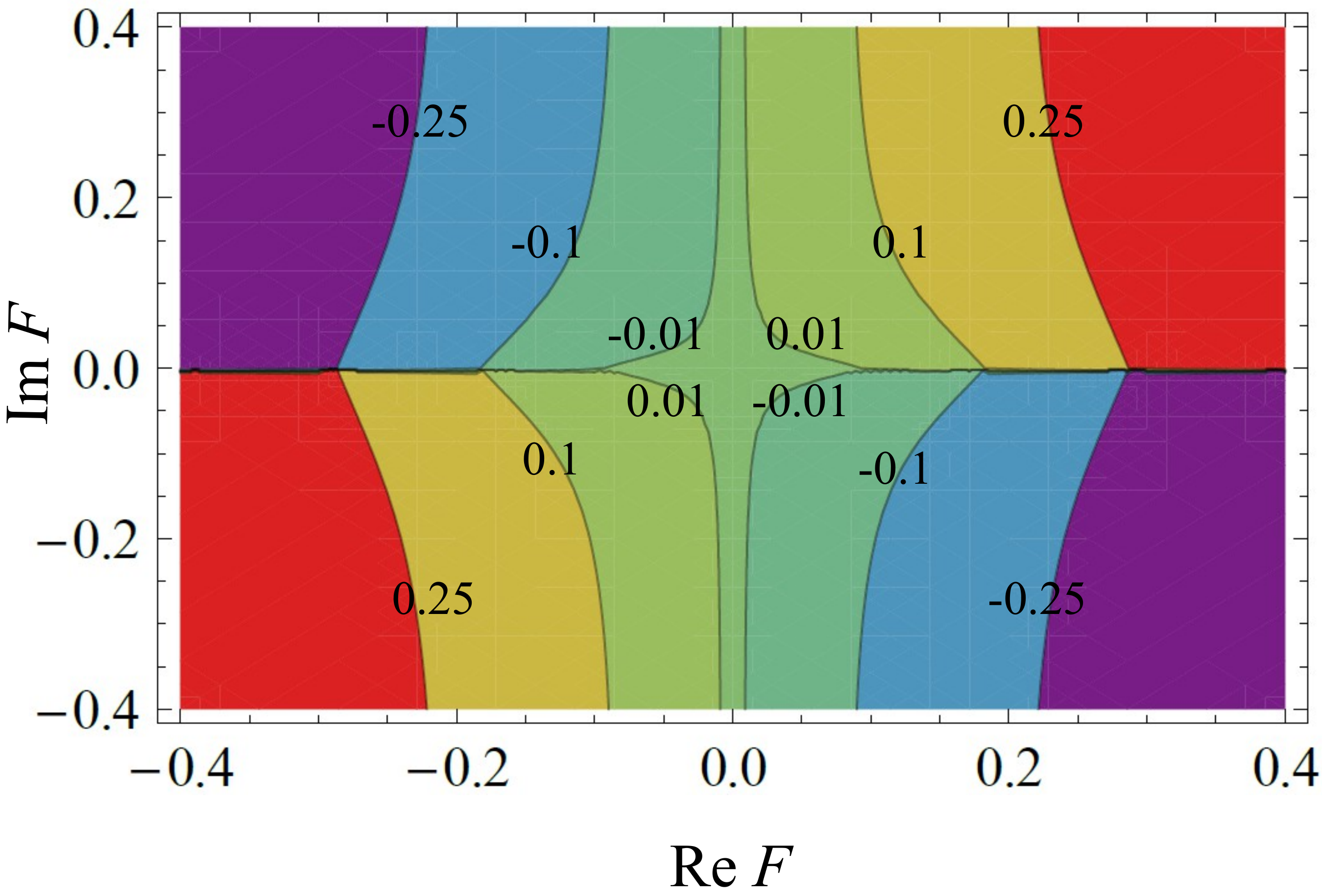}
\caption{(Color online) Imaginary part of $_2F_1(h_1,h_2,h_3,h_4 f)$ calculated for the Stark Hamiltonian in the complex $F$-plane. The built-in branch cut extends from $ (h_4)^{-1}$ to $\infty$, and is essential to obtain $\Gamma(F) \neq 0$. The $h_i$ are determined from the first four coefficients of the perturbation expansion. These yield $h_4\approx 164961$.}
\label{fig:fig3}
\end{figure}

It is easy to understand why the hypergeometric and Borel-hypergeometric method dramatically outperform the traditional Borel-Pad\'e method~\cite{Caliceti2007}. To obtain $\Gamma(F) \neq 0$ one needs approximants with a branch cut in the complex $F$ plane with branch points at $F=0$ and $F=\pm \infty$~\cite{Bender1969,Reinhardt1982}. Pad\'e approximants typically have both poles and zeroes on the real $F$ axis, thereby lacking the correct analytic structure of $E(F)$ --essential for rapid convergence. The function $_2 F_1(h_1,h_2,h_3;h_4 F)$ has a branch cut running from $h_4 F=1$ to $h_4 F = \infty$. When calculating $\{h_i\}$ from the low-order coefficients $e_1, \ldots, e_4$, we typically obtain a large value for $h_4$, thus mimicking the correct branch cut structure in $E(F)$, as  illustrated in Fig.~\ref{fig:fig3}. 

Our findings have interesting physical implications. At the end of  Ref.~\cite{Dyson1952}, Dyson wonders about the possibility that a series with zero radius of convergence might contain all that there is to know in a quantum system, arguing that if this were the case then an extension of QED would be needed.  In the three examples considered in our study,  the  ``extension'' needed consists in supplementing the low-order information with an AC function able to mimic the branch cut structure required to obtain $\Gamma(F) \neq 0$ ~\cite{Reinhardt1982, Dunne2002} .

We speculate that the approximations developed here have potential applications in nonequilibrium many-body perturbation theory~\cite{Stefanucci2013} for condensed matter systems.  For example, there is a habit of partially resumming classes of Feynman diagrams for
such problems, as exemplified by the self-consistent Born approximation (SCBA)~\cite{Frederiksen2007,Mahfouzi2013} for electron-boson interacting systems
or the self-consistent $GW$ approximation~\cite{Spataru2009} for electron-electron interacting systems out of equilibrium. These are both self-consistent  in order to conserve charge current~\cite{Stefanucci2013}, and can therefore be viewed as ``infinite order'' approximations. However, they are  exact only to first order because, to second order, SCBA misses the polarization bubble diagram~\cite{Mahfouzi2013} and the first-order vertex correction~\cite{Dash2011} to the Fock diagram, while the self-consistent $GW$ approximation accounts for the former but also misses the latter. Thus, such resummation schemes are uncontrolled because an error is ``summed'' to all orders (starting from second order)~\cite{Mera2013}. An alternative could be to build the exact diagrammatic series---including vertex corrections---at low orders, and then use a suitable AC technique. The hypergeometric and Borel-hypergeometric techniques proposed here look promising in this respect.

In conclusion, by analogy with traditional Pad\'e and Borel-Pad\'e techniques we have developed a fourth-order hypergeometric approximant and its natural Borel extension. We demonstrate the effectiveness of this technique by summing the perturbative expansions with zero radius of convergence for three well-known examples from single-particle quantum mechanics, obtaining excellent approximations to their perturbed ground state eigenenergies, even for rather large values of the perturbation strength. The imaginary part of the perturbed eigenvalue is obtained with good precision using only fourth-order perturbation theory, thereby evading the calculation of a large number of coefficients in perturbation theory. Our results show that  nonperturbative physics can be obtained from the low-order coefficients of a divergent perturbation series, as long as a carefully tailored analytic continuation technique is implemented.

\begin{acknowledgments}
H. M. and B. K. N. were supported by NSF under Grant No. ECCS 1202069. T. G. P. acknowledges funding for the QUSCOPE center by the Villum foundation.
\end{acknowledgments}

\end{document}